\documentclass{article}

\usepackage{arxiv}

\usepackage[utf8]{inputenc} % allow utf-8 input
\usepackage[T1]{fontenc}    % use 8-bit T1 fonts
\usepackage{hyperref}       % hyperlinks
\usepackage{url}            % simple URL typesetting
\usepackage{booktabs}       % professional-quality tables
\usepackage{amsfonts}       % blackboard math symbols
\usepackage{nicefrac}       % compact symbols for 1/2, etc.
\usepackage{microtype}      % microtypography
\usepackage{lipsum}		% Can be removed after putting your text content
\usepackage{graphicx}
\usepackage[square,numbers]{natbib}
\usepackage{doi}

\usepackage[export]{adjustbox}
\usepackage{tablefootnote}
\usepackage{xcolor}
\usepackage{graphicx}
\usepackage{cuted}
\usepackage{float}
\usepackage{graphicx}
\usepackage{caption}
\usepackage{comment}
\usepackage{amsmath}
\definecolor{color}{RGB}{25,25,112}
\definecolor{negro}{RGB}{0,0,0}
\definecolor{colorurl}{RGB}{25,25,112}
\usepackage{orcidlink}
\usepackage{amssymb}
\usepackage{tikz}
\usepackage{tikz-3dplot}
\usetikzlibrary{patterns}

\newcommand{\vect}[1]{\boldsymbol{#1}}

\title{Exciton Localization in Two-Dimensional Semiconductors Through Modification of the Dielectric Environment}

\author{\href{https://orcid.org/0000-0000-0000}{\includegraphics[scale=0.06]{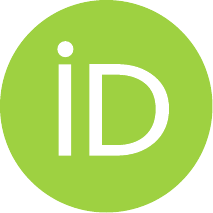}\hspace{1mm} Kelly Y. Muñoz-Gómez}  \href{https://orcid.org/0000-0003-1575-6583}{\includegraphics[scale=0.06]{orcid.pdf}\hspace{1mm} Hanz Y. Ram\'irez-G\'omez} \thanks{hanz.ramirez@uptc.edu.co} \\
	Grupo de F\'isica Te\'orica y Computacional \& Grupo QUCIT, \\
	Escuela de F\'isica, Universidad Pedag\'ogica y Tecnol\'ogica de Colombia (UPTC),\\
	Tunja 150003, Boyac\'a, Colombia. \\
}

\hypersetup{
	pdftitle={Exciton Coulomb correlations in axially symmetric QDs},
	pdfsubject={q-bio.NC, q-bio.QM},
	pdfauthor={C. A. Boh{\'o}rquez-Quincos, H. Y. Ram\'irez-G\'omez},
	pdfkeywords={Excitons, Coulomb Correlations, Semiconductor Nanoparticles},
}

\begin{document}
	\maketitle
	
\begin{abstract}
Monolayer semiconductors, given their thickness at the atomic scale, present unique electrostatic environments due to the sharp interfaces between the semiconductor film and surrounding materials. These interfaces significantly impact both the quasiparticle band structure and the electrostatic interactions between charge carriers. 

A key area of interest in these materials is the behavior of bound electron-hole pairs (excitons) within the ultra-thin layer, which plays a crucial role in its optoelectronic properties.

In this work, we investigate the feasibility of generating potential traps that completely confine excitons in the thin semiconductor by engineering the surrounding dielectric environment. By evaluating the simultaneous effects on bandgap renormalization and modifications to the strength of the electron-hole Coulomb-interaction, both associated to the modulation of the screening by the materials sandwiching the monolayer, we anticipate the existence of low-energy regions in which the localization of the exciton center of mass may be achieved. 

Our results suggest that for certain dielectric configurations, it is possible to generate complete discretization of exciton eigenenergies in the order of tens of meV. Such quantization of energy levels of two-dimensional excitons could be harnessed for applications in new-generation optoelectronic devices, which are necessary for the advancement of technologies like quantum computing and quantum communication.
\end{abstract}

	% keywords can be removed
	\keywords{Confined Excitons \and 2D Heterostructures \and Dielectric Modulation}

\section{Introduction}
Over the past two decades, two-dimensional (2D) semiconductors have garnered significant attention due to their exceptional optical and electronic properties, which differ fundamentally from those of their bulk counterparts \cite{Ayari2018, Peng2019}. These materials, characterized by their atomic-scale thickness, exhibit  charge carriers strongly confined in a plane, leading to unique excitonic effects and tunable energy band structures \cite{Control-of-gap}. A particularly compelling feature of 2D semiconductors is their ability to host excitons whose properties can be precisely controlled through external factors such as strain, electric fields, or, as explored in this study, the surrounding dielectric environment \cite{Hichri2017, Smiri2021, Cho2017}. This level of control has opened up new possibilities for applications in fields such as  quantum-light generation and single-particle optoelectronic devices.

Experimentally, 2D materials have been widely explored as light-emitting sources \cite{Pu2018}. Recent progresses in the fabrication, characterization and application of 2D heterostructures have shown how these engineered structures, typically comprising materials like graphene, hexagonal boron nitride (hBN) and transition metal dichalcogenides (TMDCs), unlock new properties and applications beyond those of their individual components \cite{Wand2014, Junlei2023}. Furthermore, substantial advancements have been made in the synthesis of van der Waals heterostructures, with their applications extending to electronics, sensing, optoelectronics and energy conversion \cite{Wang2022, Nguyen2023}.

Successful trapping of neutral excitons by means of electrical gates has been recently reported \cite{electrical-trapping-1,electrical-trapping-2}. However, scalable fabrication of those devices for field-driven confinement (active) may result disadvantageous as compared to heterostructures in which geometry- and material-driven  confinement (passive) is feasible.   

It is well established that the dielectric environment surrounding a 2D semiconductor plays a critical role in modulating the Coulomb interaction between electron and hole (e-h) pairs \cite{Raja2017, Brahma2023, Ryou2016}. In homogeneous environments, the e-h interaction is primarily governed by the relationship between the dielectric constant of the semiconductor (in-plane component) and the effective dielectric constants (average between in-plane and out-of-plane components) of the surrounding materials \cite{Rytova1967, Keldysh1979, Kumagai1990, Hanamura1988}. However, the presence of non-uniform dielectric surroundings along the in-plane directions has been shown to significantly alter this interaction, resulting in phenomena such as changes in dielectric screening, local bandgap renormalization, and enhanced carrier confinement \cite{Raja2017, Gupta2017, Salzwedel2024}. These modulations are expected to affect exciton binding energies and radiative lifetimes, both of which are crucial for designing efficient optoelectronic devices.

In low-dimensional semiconductors, bandgap renormalization arises from changes in field screening across interfaces \cite{Kleinma1985, Das1990}. This effect can be significant when there is a strong contrast between the dielectric properties of the semiconductor nanostructure and the materials surrounding it.

Additionally, the strength of the electrostatic interaction between confined carriers is sensitive to the dielectric environment and can be substantially altered when the polarizabilities of the involved materials differ significantly \cite{Rodina2016, Raja2017}.

In this work, we explore the interplay between bandgap renormalization and changes in the exciton binding energy in a 2D semiconductor, specifically considering the effects of inhomogeneities in the dielectric surroundings along the in-plane directions. We begin by introducing the system under consideration and the proposed Hamiltonian used to study the impact of the modified dielectric environment. Next, we present numerical results for a monolayer surrounded by two inhomogeneous slabs. Finally, we analyze the outcomes and provide key conclusions.
	
	\section{System Hamiltonian}

\begin{figure}[H]
	\centering		\includegraphics[width=\linewidth]{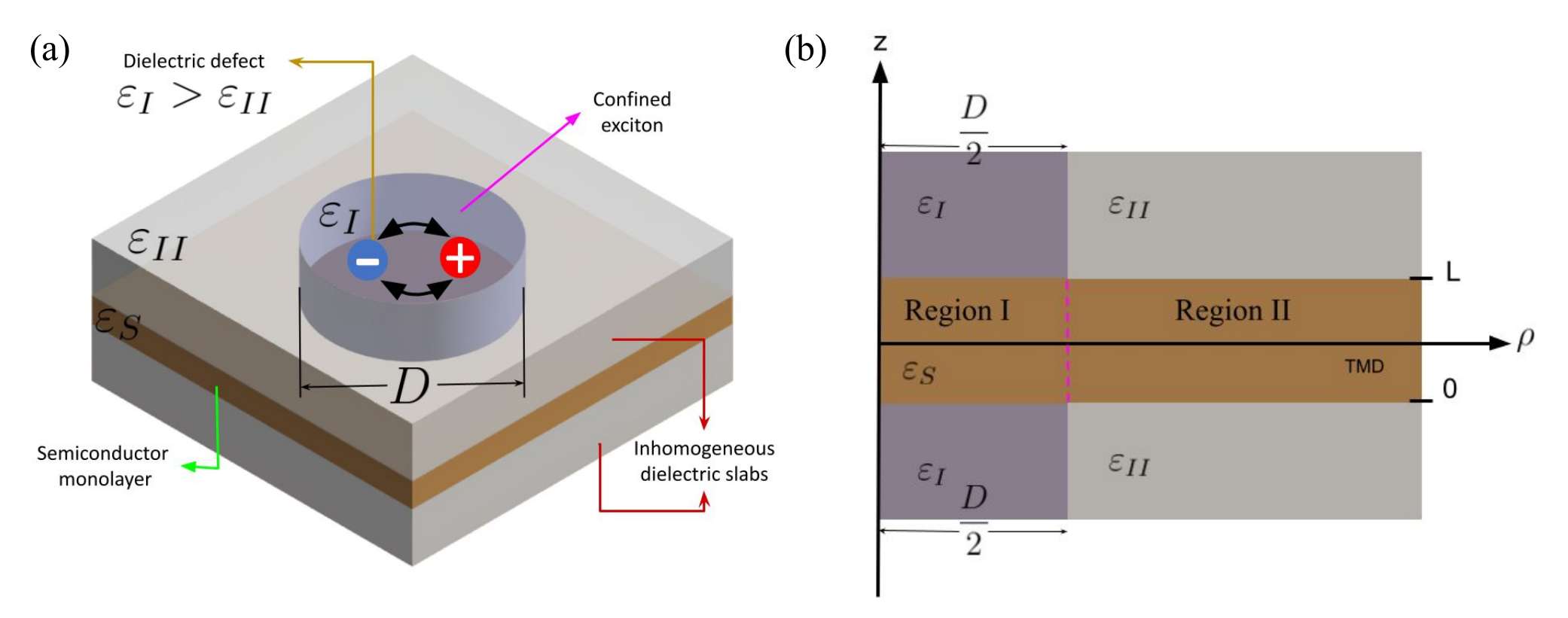}
	\caption{(a) Schematic of the studied heterostructure.  (b) Side view of the Inhomogeneous-dielectric/2D-Semiconductor/Inhomogeneous-dielectric stack. $\rho$ is the in-plane component of the relative position.}
	\label{schematics-1}
\end{figure}

The specific system under study consists of a semiconductor monolayer whose central region is sandwiched between two cylindrical rods with diameter $D$ and of a material with effective dielectric constant $\varepsilon_{I}$. Those rods are surrounded by slabs of a different material with smaller effective dielectric constant $\varepsilon_{II}$ (which could be air, i.e. $\varepsilon_{II} = 1$). This configuration is depicted in figure \ref{schematics-1}, where a perspective and a side view are provided. 

Figure \ref{schematics-1}(b) illustrates two distinct regions within the 2D semiconductor: region I ($0 \le \rho < D/2 $) and region II ($ D/2 \le \rho < \infty $), defined by the boundaries of the dielectric environment.

For e-h pairs in the 2D semiconductor, we will use the standard transformation of coordinates for the center-of-mass ($\vect{R}$) and relative ($\vect{r}$) positions \cite{Theory-centerofmass-relative-1,Theory-centerofmass-relative-2}. The origin of coordinates is taken vertically at mid-height of the 2D semiconductor, and horizontally at the center of the rods within the dielectric slabs (see figure \ref{schematics-1}). Due to the atomic-scale thickness of the semiconductor, where the excitons are confined, we set $Z=z=0$ for both the vertical e-h center-of-mass and the relative positions. In the side view shown in figure \ref{schematics-1}(b), $\rho$ stands for the magnitude of the in-plane component of the relative-position vector.

To write a Hamiltonian for this system, three assumptions are made: i) The exciton radius $r_X$ is much smaller than $D$, ii) The electron, hole and e-h center of mass are always in the same region, and iii) The electron and hole effective masses are the same in regions I and II.

The first two assumptions are based on the fact that the typical exciton radius in 2D transition metal dichalcogenides is in the order of angstroms \cite{Zhu2024}, while we will consider cylindrical rods with diameters in the order of several nanometers \cite{Tian2023}. The third assumption is reasonable because the in-plane effective masses of carriers inside the thin semiconductor are expected to mainly depend on the crystalline structure of that material, instead of being dominated by the dielectrics above and below it \cite{Shenton2018}.  

Based on the above assumptions, we describe the position-represented Hamiltonian in terms of a Heaviside function $h(x)$ ($h(x)=0$ for $x < 0$ and $h(x)=1$, otherwise), according to

\begin{equation}
	\centering 
	\begin{aligned}
		H = -\frac{\hbar^{2}}{2 M} \nabla_{\mathbf{R}}^{2} -\frac{\hbar^{2}}{2 \mu} \nabla_{\mathbf{r}}^{2} +  \left[ 1 - h(R-D) \right] \times \left[ E_{B}^I + V_I^{e-h} (\vect{r}) \right]  + h(R-D) \times \left[ E_{B}^{II} + V_{II}^{e-h} (\vect{r}) \right] ,
		\label{eq:1}
	\end{aligned}
\end{equation}

where $R$ is the magnitude of the center of mass position, $\nabla_{\mathbf{R}}$ and $\nabla_{\mathbf{r}}$ are correspondingly the Laplacians respect to the center of mass and relative positions, while $M$ and $\mu$ denote the total and reduced e-h masses, respectively. $E_{B}^I$ and $V_I^{e-h} (\vect{r})$ ($E_{B}^{II}$ and $V_{II}^{e-h} (\vect{r})$) are respectively, the quasiparticle bandgap and e-h Coulomb potential in region I (region II). Along this work we will consider only the direct part of the Coulomb interaction and neglect the exchange component, since in 2D semiconductors, the typical energy scale of the former has been shown more than one order of magnitude larger than the latter \cite{Peng2019,SJ-Cheng-nanomaterials}. 

``Bandgap renormalization'' is an observed effect associated to changes in the dielectric environment \cite{Capizzi1984, Onishi2024, Sans2008, Gil2014}. Because of it, the quantities $E_{B}^{I}$ and $E_{B}^{II}$
are expected to differ in an amount whose absolute value increases as the difference between dielectric constants of the 2D semiconductor and the locally surrounding dielectric raises.   

To account for that renormalization effect, we use the model developed by Cho and Berkelbach in reference \cite{Cho2017}. This formulation allows to obtain the change in the bandgap of a 2D sample respect to that of the bulk case, in terms of the difference between the involved dielectric constants. Accordingly, the bandgap modification $\Delta E_{B}^i$ is given by 

%This provides crucial insights into how dielectric properties affect the energy levels and excitonic states in the system, which is essential for understanding exciton confinement and energy level discretization within the semiconductor layer.

%Furthermore, the bandgap renormalization due to the dielectric environment is described by Equation \ref{eq:4}.

\begin{equation}
	\Delta E_{B}^i =  \frac{e^{2}}{ \varepsilon_{S} d} \left\{ 2 \tanh^{-1}(L_{S,i}) - \ln\left(1 - L_{S,i}^{2}\right) \right\},
	\label{eq:2}
\end{equation}

where $L_{S,i}$ is a factor depending on the difference between the involved dielectric constants in each region, which reads

\begin{equation}
	L_{S,i} = \frac{\varepsilon_{S} - \varepsilon_{i}}{\varepsilon_{S} + \varepsilon_{i}},
	\label{eq:3}
\end{equation}

with $i=I,II$.

The bandgap energy for reference $E_{B}^{ref}$, is that of the sandwiched semiconductor, but in bulk form. Respect to this value, the bandgap is renormalized according to the material surrounding the 2D semiconductor. Thus, if the material above and below that film is the semiconductor itself, there is not renormalization and the bandgap would obviously remain the same. In contrast, if the effective dielectric constant of the material in region  $i$  differs from that of the semiconductor, the renormalized bandgap in that case is defined as

\begin{equation}
	E_{B}^{i} \equiv E_{B}^{ref} + \Delta E_{B}^i ,
	\label{eq:4} 
\end{equation}

may increase or decrease respect to the reference value, depending on whether such a difference ($\varepsilon_{S} - \varepsilon_{i}$) is positive ($\Delta E_{B}^i > 0$) or negative ($\Delta E_{B}^i < 0$) \cite{Control-of-gap,Gauriot2023}.

Regarding the electrostatic interaction, because of the product with a function depending on the center of mass position $\vect{R}$,  the e-h interaction potentials are not strictly separable. However, the abrupt form of the $h(R-D)$ function allows considering those potentials as depending exclusively on the magnitude of the relative position $\rho$ within each of regions I and II. 

For those potentials $V_{i}^{eh}(\rho)$ ($i=I,II$), we use the form in terms of image charges, derived in references \cite{Kumagai1990, Hanamura1988}. This method permits to effectively incorporate the influence of the inhomogeneous dielectric environment on the electron-hole Coulomb interaction, as long as the e-h pair is assumed not to be very close to the boundary $ R_x^2 + R_y^2 = D^2 $. Such electrostatic potential, can be written in SI units as

\begin{equation}
	\centering 
	\begin{aligned}
		V_{i}^{e-h}(\rho) = \frac{1}{4 \pi \varepsilon_{0}} \left[ \frac{e^{2}}{\varepsilon_{S} \rho} + 2 \sum_{n=1}^{\infty} \frac{e^{2} L_{Si}^{2n}}{\varepsilon_{S}\left\{\rho^{2} + (2nd)^{2}\right\}^{1/2}} + 2L_{Si} \sum_{n=0}^{\infty} \frac{e^{2} L_{Si}^{2n}}{\varepsilon_{S}\left\{\rho^{2} + [(2n + 1) d]^{2}\right\}^{1/2}}\right],
		\label{eq:5}
	\end{aligned}
\end{equation}

where $d$ is the effective width of the 2D semiconductor (see figure \ref{schematics-1}(b)). We will refer to this potential as ``Kumagai-Takagahara'' (KT) potential.

Such a potential incorporates the effects of both, the top and bottom dielectric interfaces in quasi 2D systems, by means of an infinite but converging series of image charges \cite{Kumagai1990, Hanamura1988}. This potential was preferred over the more commonly used approach, known as Rytova-Keldysh (RK) potential, because the KT potential describes more accurately the e-h Coulomb interactions inside a 2D semiconductor surrounded by different types of dielectric materials. For instance, the RK potential assumes the limits $\rho \gg d$ and $\varepsilon_{S} \gg \varepsilon_{i}$, while the KT potential does not, which makes the latter  applicable in a broader range of dielectric environments. Furthermore, it has been reported that the RK potential introduces some spurious effects due to its logarithmic asymptotic behavior for very short distances  \cite{Cho2017,Dinh2018}.

Equations (\ref{eq:1}) through (\ref{eq:4}) provide a complete framework to study the net energetic effect produced by simultaneous changes in the bandgap and in the e-h electrostatic interaction, at the two different regions defined within the 2D semiconductor. Then, we can estimate the magnitude of the energy jump along the in-plane directions, that is expected to be generated because of the inhomogeneity of the slabs by which the semiconductor is sandwiched.

It is worth noting that the bandgap renormalization and the electrostatic modulation exhibit agreeing trends. i.e. for any configuration in which the bandgap increases, the e-h Coulomb potential is strengthened (its absolute value is enlarged). Similarly, in those configurations in which the bandgap decreases, the e-h Coulomb attraction is weakened.

\begin{figure}[H]
	\centering
	\includegraphics[width=\linewidth]{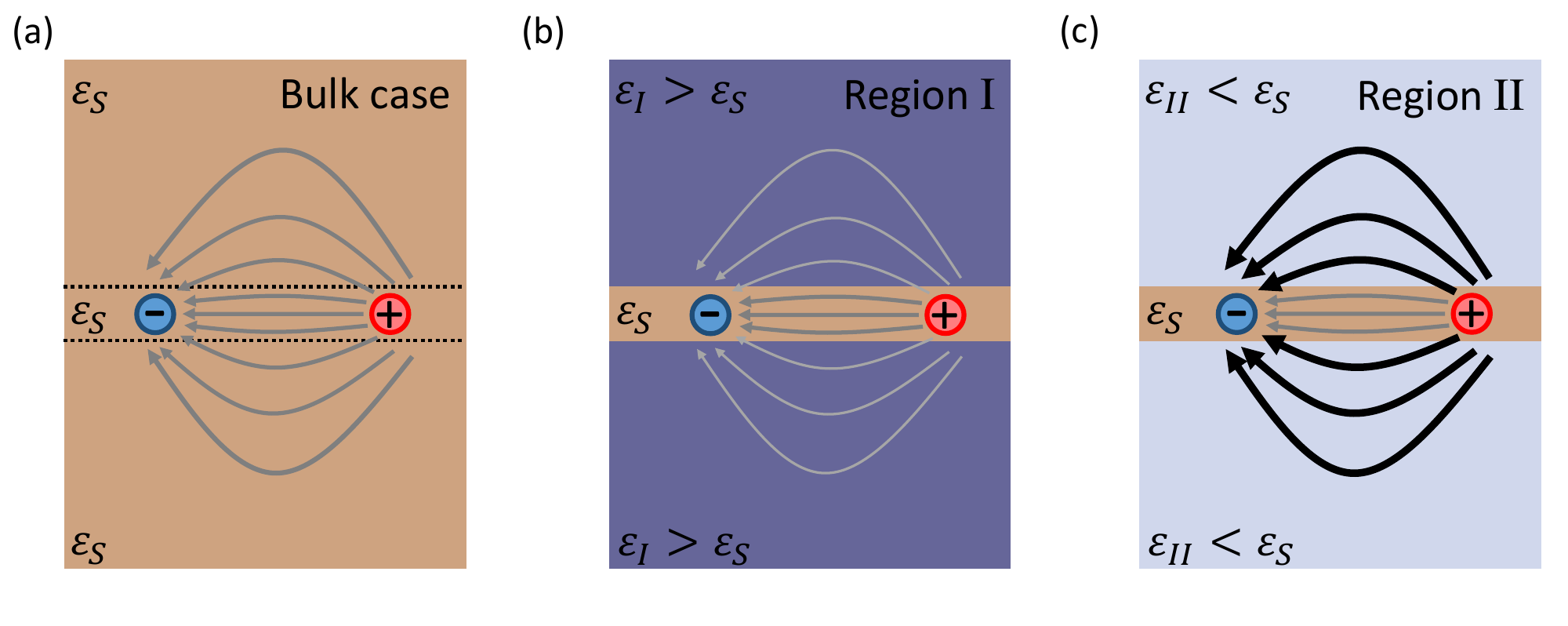}
	\caption{Depiction of the change in the electrostatic interaction depending on the dielectric surrounding. (a) A bulk sample, in which the boundaries of what would be the thin layer are represented by dashed lines. (b) The case in which the surrounding material has a larger dielectric constant than the semiconductor itself (region I). The electrostatic interaction is weakened. (c) As in (b) but the dielectric constant of the surrounding material is smaller (region II). In this case, the electrostatic interaction is strengthened.}
	\label{schematics-2}
\end{figure}

The physical explanation for this behavior is depicted in figure \ref{schematics-2}. In the illustrated case (an interacting e-h pair), when the surrounding material is more polarizable than the semiconductor itself (figure \ref{schematics-2}(b)), the attractive electrostatic interaction between the carriers is weakened in comparison to the bulk case (figure \ref{schematics-2}(a)). On the contrary, if the surrounding material is less polarizable than the semiconductor, the screening by the surrounding material is more effective and then the interaction is strengthened (figure \ref{schematics-2}(c)). Analogously happens with the repulsive interaction between electrons, which contributes to the bandgap size via the self-energy generated by the electron density in presence of the dielectric interfaces \cite{Lebens2016, Ugeda2014}.

Hence, for a region in which the 2D semiconductor is sandwiched by a material with effective dielectric constant smaller (bigger) than that of the semiconductor itself, the bandgap is expected to increase (decrease) as well as the magnitude of the exciton binding energy. Nevertheless, the energy variations associated to both of these effects are not necessarily equal. Then, there can be a net effect on the ground energy in each region.

According to the Hamiltonian in equation (\ref{eq:1}), there will be two different energies for the ground state, depending on the position of the center of mass (whether in region I or in region II). If the ground energy for region I is lower than that for region II, region I would become a confinement area for excitons within the semiconductor (a well-like effective potential $V^{eff}(R)$ for the exciton center of mass). Otherwise, region I would turn into a region where the excitons are prevented to be located (a step-like effective potential $V^{eff}(R)$ for the exciton center of mass). 

The ground energy for region $i$ can be obtained from

\begin{equation}
	E_{G}^{i} = E_{B}^{i} + E_0^{i},
	\label{eq:6}
\end{equation}

where $E_0^{i}$ ($<0$) is the energy of the ground state of the exciton relative part, calculated by solving 

\begin{equation}
	-\frac{\hbar^{2}}{2 \mu} \nabla_{\mathbf{\rho}}^{2} + V_i^{e-h} (\vect{\rho}) \Psi_0^i (\vect{\rho}) = E_{0}^{i} \Psi_0^i (\vect{\rho}).
	\label{eq:7}
\end{equation}

\begin{figure}[H]
	\centering
	\includegraphics[width=\linewidth]{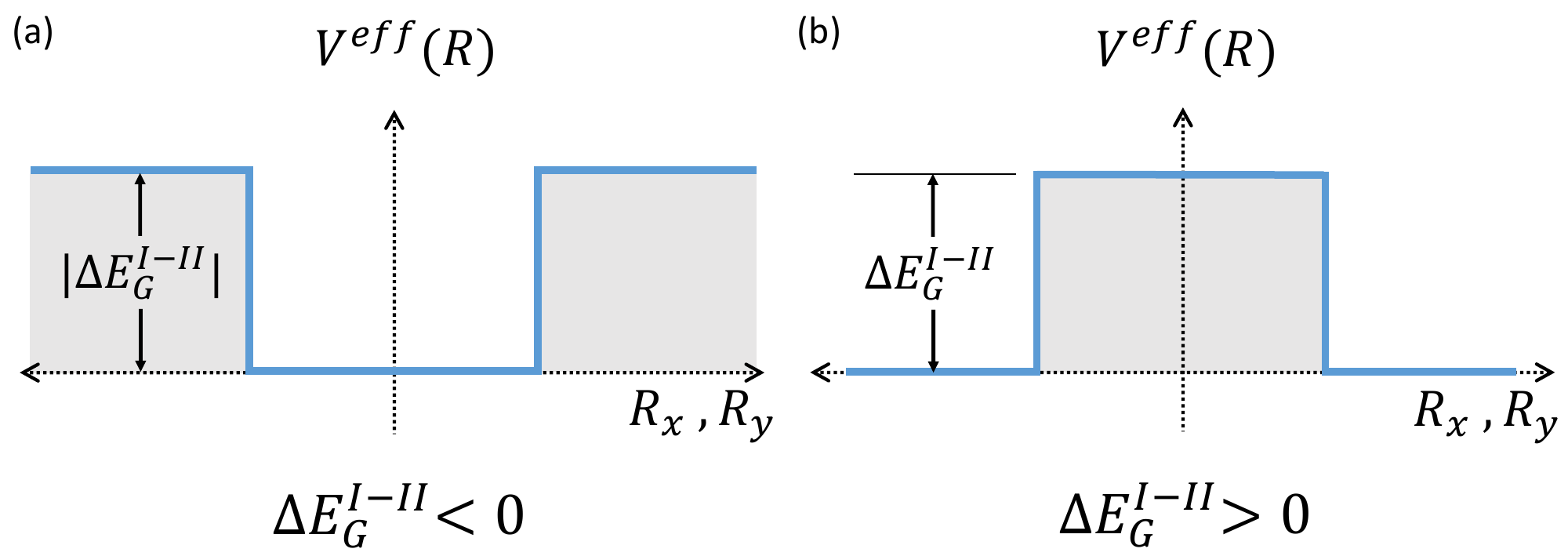}
	\caption{(a) Effective potential well for the exciton center of mass, in the case $\Delta E_{G}^{I-II} < 0$ (b) Effective potential step for the exciton center of mass in the case $\Delta E_{G}^{I-II} > 0$}
	\label{schematics-3}
\end{figure}

Then, the difference between the ground energies in region I and region II ($\Delta E_{G}^{I-II} = E_{G}^{I} - E_{G}^{II}$), can be calculated. If such a difference is positive (negative), it is considered as the effective height (depth) of a potential step (well) ``felt'' by the exciton center of mass. This is schematized in figure \ref{schematics-3}.  

\section{Numerical Results}

In this section, we study numerically three distinct configurations, corresponding to the 2D semiconductor sandwiched by three different materials in region I. In all the cases, the surrounding material in region II is taken as air ($\varepsilon_{II} = 1$), and for the in-plane component of the dielectric constant of the ultra-thin semiconductor (assumed as WS$_2$) $\varepsilon_{S} = 14$ was used. This as a representative value within the range reported for TMDC monolayers \cite{GW-TMDCs,Laturia2018}. Hence, the difference among the three studied cases is the effective dielectric constant $\varepsilon_{I}$ of the material above and below region I (see figure \ref{schematics-1}(b)).

In Case 1, region I is assumed sandwiched by a material with effective dielectric constant ($\varepsilon_{I} = 5 < \varepsilon_{S}$), i.e., a value approximated to that of Hexagonal Boron Nitride (hBN), which is a well known and widely used encapsulating insulator \cite{Knobloch2021, Laturia2018}. In Case 2, the assumed system is a semiconductor rod above y below the semiconductor monolayer ($\varepsilon_{I} = \varepsilon_{S} = 14$). This is an idealized case, useful as reference for the values of the quasiparticle bandgap and exciton binding energy related to the conventional Coulomb potential (figure \ref{schematics-2}(a), and first of the three terms in equation (\ref{eq:5})). Finally, in Case 3, region I is assumed surrounded by a material with effective dielectric constant ($\varepsilon_{I} = 25$), representing Hafnium(IV) oxide ($\text{HfO}_{2}$), i.e. a highly polarizable material  \cite{robertson2004, Liu2023}. The values of the dielectric constants used in the calculations are summarized in Table \ref{tab:1}.

\begin{table}[htbp]
	\centering
	\begin{tabular}{cccc}
		\textbf{Dielectric constant} & \textbf{Case 1} & \textbf{Case 2} & \textbf{Case 3} \\ \hline
		$\varepsilon_{S}$ & 14 & 14 & 14 \\
		$\varepsilon_{I}$ & 5 & 14 & 25 \\
		$\varepsilon_{II}$ & 1 & 1 & 1 \\ 
	\end{tabular}
	\caption{Dielectric constants used in the numerical simulations.}
	\label{tab:1}
\end{table}

Once these values are defined, we can proceed to obtain $\Delta E_{G}^{I-II}$ from equations (\ref{eq:2}), (\ref{eq:4}), and (\ref{eq:6}), for the three considered scenarios. To do that, it is necessary to solve numerically equation (\ref{eq:6}), which we do within the finite-elements framework, by means of the software COMSOL Multiphysics \cite{comsol,own-comsol-2-electrons,QD-molecules-2014,own-comsol-metallic-particles}. 

To deal with the infinite series included in the KT potential, we carried out a convergence study that allowed us to truncate such a series after the first twenty terms.   

The energy values obtained from our calculations are shown in Table \ref{tab:2}, where the reference value $E_{B}^{ref}$ (the energy bandgap for bulk WS$_2$), is included for the sake of completeness. 	

\begin{table}[htbp]
	\centering
	\begin{tabular}{cccc}
		\textbf{Energy (eV)} & \textbf{Case 1} & \textbf{Case 2} & \textbf{Case 3} \\ \hline
		$E_{B}^{ref}$ & 1.350 & 1.350 & 1.350 \\
		$E_{B}^{I}$ & 1.570 & 1.350 & 1.265 \\
		$E_{B}^{II}$ & 2.040 & 2.040 & 2.040 \\
		$E_{0}^{I}$ & -0.237 & -0.092 & -0.049 \\
		$E_0^{II}$ & -0.637 & -0.637 & -0.637  \\
		$\Delta E_{G}^{I-II}$ & 0.071 & 0.146 & 0.188 \\
	\end{tabular}
	\caption{Energy values obtained for the three scenarios under study (eV). The value $E_{B}^{ref}$ is taken for bulk WS$_2$ from references \cite{WS2-gap-1,WS2-gap-2}}
	\label{tab:2}
\end{table}

\begin{figure}[htbp]
	\centering
	\includegraphics[width=1\linewidth]{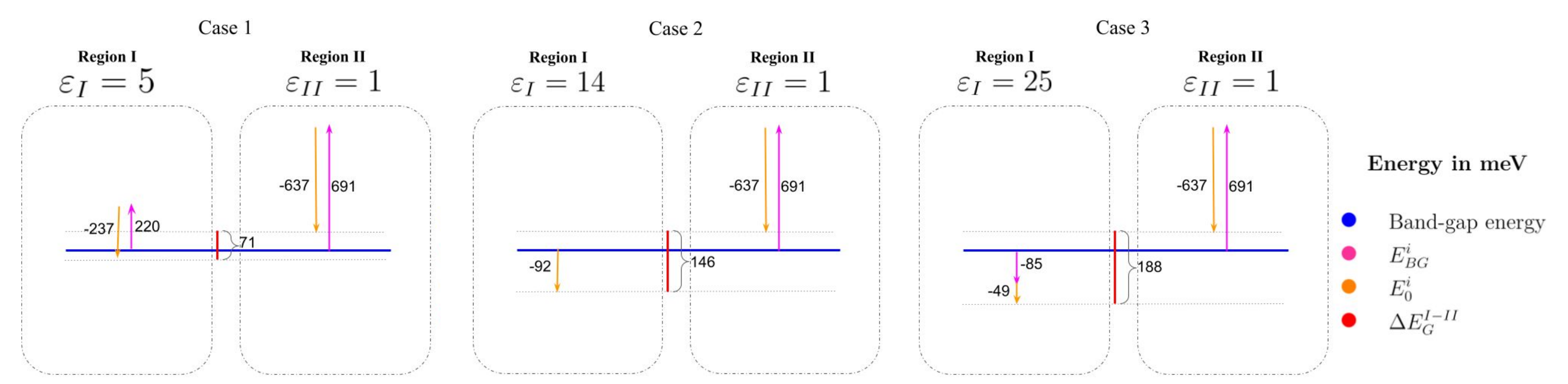}
	\caption{Visualization of the competing effects from bandgap renormalization and modulated Coulomb interaction. $E_{B}^{ref}$ (blue line), $\Delta E_{B}^{i}$ (magenta line), $E_{0}^{i}$ (orange line)  and $\Delta E_{G}^{I-II}$ (red line). }
	\label{fig:4}
\end{figure}

In figure \ref{fig:4}, the information presented in Table \ref{tab:2} is graphically illustrated. There, the competing effects from bandgap renormalization and Coulomb-interaction modulation can be visualized. The blue horizontal line indicates the bandgap energy when there is only semiconductor material, with a dielectric constant \( \varepsilon_{S} \). The magenta vertical lines represent the bandgap renormalization energies \( E_{B}^{i} \) calculated using equation (\ref{eq:2}), while the orange vertical lines show in each region the ground-state energies for the relative exciton part \( E_{0}^{i} \), calculated from equation (\ref{eq:5}) by using the KT potential. The red lines highlight the energy difference between the two effects for each studied case ($\Delta E_{G}^{I-II}$).

The red lines in figure \ref{fig:4}, highlight the net effect of the changes in dielectric environment between region I and II. For the three considered scenarios, the observed behavior is similar: a potential well is formed for the exciton center of mass, with an effective energy barrier ($\Delta E_{G}^{I-II}$) in the order of few hundreds of meV. From this, it can be expected that in any of these cases, excitons are much more likely to be located in region I than in region II (figure \ref{schematics-3}(a)).

If the effective dielectric constants of the materials defining regions I and II were exchanged, a potential step around the coordinate origin would be formed in all three cases (figure \ref{schematics-3}(b)), and excitons would be repelled from region II, generating an exciton-free region within the 2D semiconductor. 

The potential profiles in figures \ref{schematics-3}(a) and \ref{schematics-3}(b), are expected to be softened if the model would include microscopic details on the charge density induced in the dielectric interfaces. The magnitude of this effect on the energy discretization will directly depend on the softening length, which in turn will depend inversely on the difference between dielectric constants, i.e. the larger the latter the shorter the former. In this study, a negligible softening length is taken, since the considered materials have a noticeable mismatch in their dielectric constants.

To estimate for a given $D$, the energy scale in which the center-of-mass eigenenergies would be discretized as a consequence of the effectively generated potential well, we need to solve the eigenvalue equation 

\begin{equation}
	-\frac{\hbar^{2}}{2 M} \nabla_{\mathbf{R}}^{2} + V^{eff} (\vect{R}) \phi_n (\vect{R}) = E_{0}^{C} \phi_n (\vect{R}),
	\label{eq:6}
\end{equation}

where $V^{eff} (\vect{R}) = |\Delta E_{G}^{I-II}| \times h( R - D)$ ($V^{eff} (\vect{R}) = 0$ if $R < D$ and $V^{eff} (\vect{R}) = |\Delta E_{G}^{I-II}|$ otherwise), and $E_{0}^{C}$ are the discretized energies related to the center-of-mass confinement. To obtain those values, we suppose holes in the dielectric slabs of diameter $D=5$ nm, and again find numerical solutions by means of the finite element method, as implemented in references \cite{Ramirez-2006,Ramirez-2007,Rodriguez2008,Comsol-Delaware}. 

Although the fabrication of dielectric rods with such small diameter is indeed challenging, the progresses reported on patterned etching at the few-nanometer level, suggest that this type of heterostructures may be actually produced \cite{nanofabrication-1,nanofabrication-2}

The obtained values for the first three eigenenergies in each of the studied configurations are shown in Table \ref{tab:3}. $E_{1}^C$ and $E_{2}^C$ are degenerated in all cases because of the axial symmetry of the system.

\begin{table}[htbp]
	\centering
	\begin{tabular}{cccc}
		\textbf{Energy levels} & \textbf{Case 1} & \textbf{Case 2} & \textbf{Case 3} \\ \hline
		$E_{0}^C$ & 16.48 & 18.92 & 19.65  \\
		$E_{1,2}^C$ & 40.55 & 47.51 & 49.52  \\
		$E_{3}^C$ & 68.76 & 84.16 & 88.16  \\
	\end{tabular}
	\caption{Discretized center-of-mass energies (meV). $E_{1,2}$ are degenerated in each case.}
	\label{tab:3}
\end{table}

The values presented in Table \ref{tab:3}, show that the energy discretization would be very similar in any of the three considered cases, being the exciton localization almost insensitive to the specific type of dielectric used in region I. 

The obtained energy values suggest a significant discretization of levels, big enough for eventual experimental observation, even without the use of cryogenic setups. Further work needs to be done, for finding additional configurations that would lead to similar discretization but with larger inhomogeneities in the dielectric environment, so that the experimental realization is facilitated. 

Such a discretization, of the order of tens of meV, seems advantageous for diverse applications that would benefit from combining the attractive properties of 2D materials and zero-dimensional nanostructures. In particular, regarding single photon emission, to evaluate the feasibility of using the studied heterostructures for such application, we estimate the magnitude of the corresponding exciton-exciton interaction by means of the formulation presented in reference \cite{exciton-exciton-interaction}. The obtained interaction energy is around 100 meV, which is one order of magnitude larger than the linewidth of the exciton emission in monolayer TMDCs (typically 5-10 meV) \cite{exciton-linewidth}. This makes exciton blockade clearly achievable within the confinement region, favoring photon antibunching.   

This dielectric-driven effective trapping suggested by our calculations, is consistent with observations of funneling of neutral excitons reported in \cite{Gauriot2023,Exciton-funneling-2022}  

It is important to mention that the presented formulation allows to describe systems with different regions in which the dielectric surrounding is the same at the top and bottom of the 2D semiconductor, but cannot be applied in vertically asymmetric cases. Hopefully, an extended framework useful for studying a wider range of heterostructures, will be available in a further work.

\section{Conclusions}

We have studied the possibility of localizing excitons in 2D semiconductors, by modifying the dielectric environment. The obtained numerical results demonstrated that inhomogeneities in the dielectric environment may result in complete discretization of exciton energies. 

Such quantization of energy levels, indicates the possibility of creating quantum dots in an ultra-thin semiconductor film. This offers valuable insight into how exciton confinement can be achieved through dielectric modulation.

These findings are expected to contribute to advancements in the development of artificial atoms with tunable properties, which are promising for the fabrication of high-performance quantum light sources, that are in turn, essential for the development of photon-based quantum technologies, such as quantum communication and quantum computing.

	% Acknowledgements
\medskip
\textbf{Funding}\\ 
%\par %delete if not applicable))
This research was funded by Sistema General de Regalías Colombia, through grant number BPIN2021000100191, and by the Research Division of UPTC through project SGI-3378.\\

\textbf{Acknowledgments}\par
The authors thank Prof. Shun-Jen Cheng and Dr. Jhen-Dong Lin from NYCU university in Taiwan, for inspiring discussions.

\bibliographystyle{ieeetr}
\bibliography{biblio-exciton-localization-fv.bib}

@article{Ayari2018,
author = {Ayari, Sabrine and Smiri, Adlen and Hichri, Aïda and Jaziri, S. and Amand, Thierry},
year = {2018},
month = {03},
pages = {},
title = {Radiative lifetime of localized excitons in transition metal dichalcogenides},
volume = {98},
journal = {Physical review. B, Condensed matter},
doi = {10.1103/PhysRevB.98.205430}
}

@article{Peng2019,
author = {Peng, Guan-Hao and Lo, Ping-Yuan and Li, Wei-Hua and Huang, Yan-Chen and Chen, Yan-Hong and Lee, Chi-Hsuan and Yang, Chih-Kai and Cheng, Shun-Jen},
year = {2019},
month = {03},
pages = {},
title = {Distinctive Signatures of the Spin- and Momentum-Forbidden Dark Exciton States in the Photoluminescence of Strained WSe 2 Monolayers under Thermalization},
volume = {19},
journal = {Nano Letters},
doi = {10.1021/acs.nanolett.8b04786}
}

@article{Rytova1967,
author = {N. S. Rytova},
journal = {Proc. MSU, Phys.},
year = {1967},
month = {06},
pages = {},
title = {Screened potential of a point charge in a thin film}
}

@article{Keldysh1979,
    title = {Coulomb interaction in thin semiconductor and semimetal films},
    author = {L. V. Keldysh},
    journal = {Soviet Journal of Experimental and Theoretical Physics Letters},
    volume = {29},
    issue = {11},
    pages = {716},
    year = {1979},
    doi = {},
    url = {http://jetpletters.ru/ps/0/article_22207.shtml},
}

@article{Kumagai1990,
author = {Kumagai, Masami and Takagahara, Toshihide},
year = {1990},
month = {01},
pages = {12359-12381},
title = {Excitonic and nonlinear-optical properties of dielectric quantum-well structures},
volume = {40},
journal = {Physical review. B, Condensed matter},
doi = {10.1103/PhysRevB.40.12359}
}

@article{Hanamura1988,
author = {Hanamura, Eiichi and Nagaosa, Naoto and Kumagai, Masami and Takagahara, Toshihide},
year = {1988},
month = {12},
pages = {255-258},
title = {Quantum wells with enhanced exciton effects and optical non-linearity},
volume = {1},
journal = {Materials Science and Engineering B-advanced Functional Solid-state Materials},
doi = {10.1016/0921-5107(88)90006-2}
}

@article{Cho2017,
author = {Cho, Yeongsu and Berkelbach, Timothy},
year = {2017},
month = {09},
pages = {},
title = {Environmentally-Sensitive Theory of Electronic and Optical Transitions in Atomically-Thin Semiconductors},
volume = {97},
journal = {Physical Review B},
doi = {10.1103/PhysRevB.97.041409}
}

@article{Dinh2018,
author = {Dinh Van, Tuan and Yang, Min and Dery, Hanan},
year = {2018},
month = {09},
pages = {},
title = {The Coulomb interaction in monolayer transition-metal dichalcogenides},
volume = {98},
journal = {Physical Review B},
doi = {10.1103/PhysRevB.98.125308}
}

@article{Gauriot2023,
author = {Gauriot, Nicolas and Ashoka, Arjun and Lim, Juhwan and See, Soo and Sung, Jooyoung and Ananth, Akshay Rao},
year = {2023},
month = {12},
pages = {},
title = {Direct Imaging of Carrier Funneling in a Dielectric Engineered 2D Semiconductor},
volume = {18},
journal = {ACS Nano},
doi = {10.1021/acsnano.3c05957}
}

@article{Hichri2017,
author = {Hichri, Aïda and Amara, Imen and Ayari, Sabrine and Jaziri, S.},
year = {2017},
month = {06},
pages = {235702},
title = {Exciton center-of-mass localization and dielectric environment effect in monolayer WS 2},
volume = {121},
journal = {Journal of Applied Physics},
doi = {10.1063/1.4984790}
}

@article{Smiri2021,
author = {Smiri, Adlen and Amand, Thierry and Jaziri, S.},
year = {2021},
month = {03},
pages = {},
journal = {The Journal of chemical physics},
title = {Optical properties of exciton in two-dimensional transition metal dichalcogenide nanobubbles},
doi = {10.48550/arXiv.2005.03166}
}

@article{Gupta2017,
author = {Gupta, Garima and Kallatt, Sangeeth and Majumdar, Kausik},
year = {2017},
month = {03},
pages = {},
title = {Direct observation of giant binding energy modulation of exciton complexes in monolayer MoSe 2},
volume = {96},
journal = {Physical Review B},
doi = {10.1103/PhysRevB.96.081403}
}

@article{Raja2017,
author = {Raja, Archana and Chaves, Andrey and Yu, Jaeeun and Arefe, Ghidewon and Hill, Heather and Rigosi, Albert and Berkelbach, Timothy and Nagler, Philipp and Schüller, Christian and Korn, Tobias and Nuckolls, Colin and Hone, James and Brus, Louis and Heinz, Tony and Reichman, David and Chernikov, Alexey},
year = {2017},
month = {05},
pages = {15251},
title = {Coulomb engineering of the bandgap and excitons in two-dimensional materials},
volume = {8},
journal = {Nature Communications},
doi = {10.1038/ncomms15251}
}

@article{Salzwedel2024,
author = {Salzwedel, Robert and Greten, Lara and Schmidt, Stefan and Hughes, Stephen and Knorr, Andreas and Selig, Malte},
year = {2024},
month = {01},
pages = {},
title = {Spatial exciton localization at interfaces of metal nanoparticles and atomically thin semiconductors},
volume = {109},
journal = {Physical Review B},
doi = {10.1103/PhysRevB.109.035309}
}

@misc{comsol,
  author={COMSOL Inc.},
  title={COMSOL},
  year={2020},
  url={http://www.comsol.com/products/multiphysics/},
}

@article{Pu2018,
author = {Pu, Jiang and Takenobu, Taishi},
year = {2018},
month = {06},
pages = {1707627},
title = {Monolayer Transition Metal Dichalcogenides as Light Sources},
volume = {30},
journal = {Advanced Materials},
doi = {10.1002/adma.201707627}
}

@article{Wand2014,
author = {Wang, Hong and Liu, Fucai and Fang, Zheyu and Zhou, Wu and Liu, Zheng},
year = {2014},
month = {08},
pages = {},
title = {Two-dimensional heterostructures: Fabrication, characterization, and application},
volume = {6},
journal = {Nanoscale},
doi = {10.1039/C4NR03435J}
}

@article{Junlei2023,
author = {Junlei, Qi and Wu, Zongxiao and Wang, Wenbin and Bao, Kai and Wang, Lingzhi and Wu, Jingkun and Ke, Chengxuan and Xu, Yue and He, Qiyuan},
year = {2023},
month = {04},
pages = {},
title = {Fabrication and Applications of van der Waals heterostructures},
volume = {5},
journal = {International Journal of Extreme Manufacturing},
doi = {10.1088/2631-7990/acc8a1}
}

@article{Nguyen2023,
author = {Nguyen, Van Huy and Kim, Minwook and Nguyen, Cao Thang and Suleman, Muhammad and Cong, Nguyen and Nasir, Naila and Rehman, Malik and Park, Hyun and Lee, Sohee and Kim, Sung and Kumar, Sunil and Seo, Yongho},
year = {2023},
month = {08},
pages = {158186},
title = {Fast fabrication technique for high-quality van der Waals heterostructures using inert shielding gas environment},
volume = {639},
journal = {Applied Surface Science},
doi = {10.1016/j.apsusc.2023.158186}
}

@article{Wang2022,
author = {Wang, Zenghui and Xu, Bo and Pei, Shenghai and Zhu, Jiankai and Wen, Ting and Jiao, Chenyin and Li, Jing and Zhang, Maodi and Xia, Juan},
year = {2022},
month = {09},
pages = {},
title = {Recent progress in 2D van der Waals heterostructures: fabrication, properties, and applications},
volume = {65},
journal = {Science China Information Sciences},
doi = {10.1007/s11432-021-3432-6}
}

@article{Das1990,
author = {Das Sarma, Sankar and Jalabert, R and Yang, S},
year = {1990},
month = {05},
pages = {8288-8294},
title = {Band-gap renormalization in semiconductor quantum wells.},
volume = {41},
journal = {Physical review. B, Condensed matter}
}

@article{Kleinma1985,
author = {Kleinman, DA and Miller, R.},
year = {1985},
month = {09},
pages = {2266-2272},
title = {Band-gap renormalization in semiconductor quantum wells containing carriers},
volume = {32},
journal = {Physical review. B, Condensed matter},
doi = {10.1103/PhysRevB.32.2266}
}

@article{Rodina2016,
author = {Rodina, Anna and Efros, Al},
year = {2016},
month = {03},
pages = {554-566},
title = {Effect of dielectric confinement on optical properties of colloidal nanostructures},
volume = {122},
journal = {Journal of Experimental and Theoretical Physics},
doi = {10.1134/S1063776116030183}
}

@article{Zhu2024,
author = {Zhu, Tao and Zheng, Chenhang and Xu, Lei and Yang, Ming},
year = {2024},
month = {10},
pages = {},
title = {Exciton dissociation in two-dimensional transition metal dichalcogenides: Excited states and substrate effects},
volume = {110},
journal = {Physical Review B},
doi = {10.1103/PhysRevB.110.155416}
}

@article{Tian2023,
author = {Tian, Yu and Liu, Qi and Ma, Yun and Wang, Nuo and Gu, Ying},
year = {2023},
month = {10},
pages = {},
title = {Dielectric resonances of the cylindrical micro/nano cavity within epsilon-near-zero materials},
volume = {31},
journal = {Optics Express},
doi = {10.1364/OE.504233}
}

@article{Gil2014,
author = {Gil, Lyudmila and Lippi, Gian Luca},
year = {2014},
month = {11},
pages = {},
title = {Phase instabilities in semiconductor lasers: A codimension-2 analysis},
volume = {90},
journal = {Physical Review A},
doi = {10.1103/PhysRevA.90.053838}
}

@article{Sans2008,
author = {Sans, J. and Sánchez Royo, Juan Francisco and Segura, Alfredo},
year = {2008},
month = {04},
pages = {362-367},
title = {Study of the bandgap renormalization in Ga-doped ZnO films by means of optical absorption under high pressure and photoelectron spectroscopy},
volume = {43},
journal = {Superlattices and Microstructures - SUPERLATTICE MICROSTRUCT},
doi = {10.1016/j.spmi.2007.12.020}
}

@article{Capizzi1984,
author = {Capizzi, Mario and Modesti, Silvio and Frova, Andrea and Staehli, J. and Guzzi, M. and Logan, R.},
year = {1984},
month = {02},
pages = {2028-2035},
title = {Electron-hole plasma in direct-gap Ga1-xAlx As and k-selection rule},
volume = {29},
journal = {Physical Review B},
doi = {10.1103/PhysRevB.29.2028}
}

@article{Onishi2024,
author = {Onishi, Yugo and Fu, Liang},
year = {2024},
month = {10},
pages = {},
title = {Universal relation between energy gap and dielectric constant},
volume = {110},
journal = {Physical Review B},
doi = {10.1103/PhysRevB.110.155107}
}

@article{Brahma2023,
author = {Brahma, Madhuchhanda and Van de Put, Maarten and Chen, Edward and Fischetti, Massimo and Vandenberghe, William},
year = {2023},
month = {03},
pages = {},
title = {The importance of the image forces and dielectric environment in modeling contacts to two-dimensional materials},
volume = {7},
journal = {npj 2D Materials and Applications},
doi = {10.1038/s41699-023-00372-6}
}

@article{Ryou2016,
author = {Ryou, Junga and Kim, Yong-Sung and Kc, Santosh and Cho, Kyeongjae},
year = {2016},
month = {07},
pages = {29184},
title = {Monolayer MoS2 Bandgap Modulation by Dielectric Environments and Tunable Bandgap Transistors},
volume = {6},
journal = {Scientific reports},
doi = {10.1038/srep29184}
}

@article{robertson2004,
        title={ High dielectric constant oxides },
        author={ John Robertson },
        journal={ European Physical Journal-applied Physics },
        year={ 2004 },
        publisher={ EDP Sciences },
        volume={ 28 },
        pages={ 265-291 },
        number={ 3 }, 
}

@article{Liu2023,
author = {Liu, Jiangwei and Okamura, Masayuki and Mashiko, Hisanori and Imura, Masataka and Liao, Meiyong and Kikuchi, Ryosuke and Suzuka, Michio and Koide, Yasuo},
year = {2023},
month = {04},
pages = {1256},
title = {Experimental Formation and Mechanism Study for Super-High Dielectric Constant AlOx/TiOy Nanolaminates},
volume = {13},
journal = {Nanomaterials},
doi = {10.3390/nano13071256}
}

@article{Shenton2018,
author = {Shenton, Kane and Bowler, David and Cheah, Wei Li},
year = {2018},
month = {04},
pages = {},
title = {Influence of crystal structure on charge carrier effective masses in BiFeO3},
volume = {100},
journal = {Physical Review B},
doi = {10.1103/PhysRevB.100.085120}
}

@article{Laturia2018,
author = {Laturia, Akash and Van de Put, Maarten and Vandenberghe, William},
year = {2018},
month = {03},
pages = {},
title = {Dielectric properties of hexagonal boron nitride and transition metal dichalcogenides: from monolayer to bulk},
volume = {2},
journal = {npj 2D Materials and Applications},
doi = {10.1038/s41699-018-0050-x}
}

@article{Ugeda2014,
author = {Ugeda, Miguel and Bradley, Aaron and Shi, Su-Fei and Da Jornada, Felipe and Zhang, Yi and Qiu, Diana and Mo, Sung-Kwan and Hussain, Zahid and Shen, Zhi-Xun and Wang, Feng and Louie, Steven and Crommie, Michael},
year = {2014},
month = {04},
pages = {},
title = {Giant bandgap renormalization and excitonic effects in a monolayer transition metal dichalcogenide semiconductor},
volume = {13},
journal = {Nature materials},
doi = {10.1038/nmat4061}
}

@article{Lebens2016,
author = {Lebens-Higgins, Zachary and Scanlon, David and Paik, Hanjong and Sallis, Shawn and Nie, Yuefeng and Uchida, Masaki and Quackenbush, Nicholas and Wahila, Matthew and Sterbinsky, G. E and Arena, Dario and Woicik, J. C and Schlom, D. G and Piper, Louis},
year = {2016},
month = {01},
pages = {027602},
title = {Direct Observation of Electrostatically Driven Band Gap Renormalization in a Degenerate Perovskite Transparent Conducting Oxide},
volume = {116},
journal = {Phys. Rev. Lett.},
doi = {10.1103/PhysRevLett.116.027602}
}

@Article{WS2-gap-1,
	author={Kam, K. K.
	and Parkinson, B. A.},
	title={Detailed photocurrent spectroscopy of the semiconducting group VIB transition metal dichalcogenides},
	journal={The Journal of Physical Chemistry},
	year={1982},
	month={Feb},
	day={01},
	publisher={American Chemical Society},
	volume={86},
	number={4},
	pages={463-467},
	issn={0022-3654},
	doi={10.1021/j100393a010},
	url={https://doi.org/10.1021/j100393a010}
}

@article{WS2-gap-2,
	doi = {10.1149/1.2124184},
	url = {https://dx.doi.org/10.1149/1.2124184},
	year = {1982},
	month = {jul},
	publisher = {The Electrochemical Society, Inc.},
	volume = {129},
	number = {7},
	pages = {1461},
	author = {Joseph A. Baglio and Gary S. Calabrese and Emil Kamieniecki and Robert Kershaw and Clifford P. Kubiak and Antonio J. Ricco and Aaron Wold and Mark S. Wrighton and Glenn D. Zoski},
	title = {Characterization of n‐Type Semiconducting Tungsten Disulfide Photoanodes in Aqueous and Nonaqueous Electrolyte Solutions: Photo‐oxidation of Halides with High Efficiency},
	journal = {Journal of The Electrochemical Society}
}

@article{GW-TMDCs,
	title = {Theory of neutral and charged excitons in monolayer transition metal dichalcogenides},
	author = {Berkelbach, Timothy C. and Hybertsen, Mark S. and Reichman, David R.},
	journal = {Phys. Rev. B},
	volume = {88},
	issue = {4},
	pages = {045318},
	numpages = {6},
	year = {2013},
	month = {Jul},
	publisher = {American Physical Society},
	doi = {10.1103/PhysRevB.88.045318},
	url = {https://link.aps.org/doi/10.1103/PhysRevB.88.045318}
}

@article{Ramirez-2006,
	title={DC electric field effects on the electron dynamics in double rectangular quantum dots},
	author={Ramirez, HY and Camacho, AS and Lew Yan Voon, LC},
	journal={Brazilian Journal of Physics},
	volume={36},
	pages={869--873},
	year={2006},
	publisher={SciELO Brasil}
}

@article{Ramirez-2007,
	doi = {10.1088/0953-8984/19/34/346216},
	url = {https://dx.doi.org/10.1088/0953-8984/19/34/346216},
	year = {2007},
	month = {jul},
	publisher = {},
	volume = {19},
	number = {34},
	pages = {346216},
	author = {H Y Ramirez and A S Camacho and L C Lew Yan Voon},
	title = {Influence of shape and electric field on electron relaxation and coherent response in
	quantum-dot molecules},
	journal = {Journal of Physics: Condensed Matter}
}

@article{own-comsol-2-electrons,
	title = {Two interacting electrons confined in a 3D parabolic cylindrically symmetric potential, in presence of axial magnetic field: A finite element approach},
	journal = {Computer Physics Communications},
	volume = {183},
	number = {8},
	pages = {1654-1657},
	year = {2012},
	issn = {0010-4655},
	doi = {https://doi.org/10.1016/j.cpc.2012.03.002},
	url = {https://www.sciencedirect.com/science/article/pii/S0010465512001075},
	author = {Hanz Y. Ramírez and Alejandro Santana}
}

@article{own-comsol-metallic-particles,
	title = {Influence of the confinement potential on the size-dependent optical response of metallic nanometric particles},
	journal = {Computer Physics Communications},
	volume = {227},
	pages = {1-7},
	year = {2018},
	issn = {0010-4655},
	doi = {https://doi.org/10.1016/j.cpc.2018.02.012},
	url = {https://www.sciencedirect.com/science/article/pii/S0010465518300407},
	author = {Mario Zapata-Herrera and Ángela S. Camacho and Hanz Y. Ramírez}
}

@Article{SJ-Cheng-nanomaterials,
	AUTHOR = {Li, Wei-Hua and Lin, Jhen-Dong and Lo, Ping-Yuan and Peng, Guan-Hao and Hei, Ching-Yu and Chen, Shao-Yu and Cheng, Shun-Jen},
	TITLE = {The Key Role of Non-Local Screening in the Environment-Insensitive Exciton Fine Structures of Transition-Metal Dichalcogenide Monolayers},
	JOURNAL = {Nanomaterials},
	VOLUME = {13},
	YEAR = {2023},
	NUMBER = {11},
	ARTICLE-NUMBER = {1739},
	URL = {https://www.mdpi.com/2079-4991/13/11/1739},
	PubMedID = {37299642},
	ISSN = {2079-4991},
	DOI = {10.3390/nano13111739}
}

@Article{Comsol-Delaware,
	author={Welsch, Tory A.
	and Doty, Matthew F.},
	title={PbS/CdS Core/Shell Quantum Dots Designed to Enable Efficient Photon Upconversion for Solar Energy Applications},
	journal={ACS Applied Optical Materials},
	year={2024},
	month={Oct},
	day={25},
	publisher={American Chemical Society},
	volume={2},
	number={10},
	pages={2184-2195},
	doi={10.1021/acsaom.4c00340},
	url={https://doi.org/10.1021/acsaom.4c00340}
}

@article{Theory-centerofmass-relative-1,
	title = {Full-zone valley polarization landscape of finite-momentum exciton in transition metal dichalcogenide monolayers},
	author = {Lo, Ping-Yuan and Peng, Guan-Hao and Li, Wei-Hua and Yang, Yi and Cheng, Shun-Jen},
	journal = {Phys. Rev. Res.},
	volume = {3},
	issue = {4},
	pages = {043198},
	numpages = {9},
	year = {2021},
	month = {Dec},
	publisher = {American Physical Society},
	doi = {10.1103/PhysRevResearch.3.043198},
	url = {https://link.aps.org/doi/10.1103/PhysRevResearch.3.043198}
}

@article{Theory-centerofmass-relative-2,
	title = {Tailoring the superposition of finite-momentum valley exciton states in transition-metal dichalcogenide monolayers by using polarized twisted light},
	author = {Peng, Guan-Hao and Sanchez, Oscar Javier Gomez and Li, Wei-Hua and Lo, Ping-Yuan and Cheng, Shun-Jen},
	journal = {Phys. Rev. B},
	volume = {106},
	issue = {15},
	pages = {155304},
	numpages = {11},
	year = {2022},
	month = {Oct},
	publisher = {American Physical Society},
	doi = {10.1103/PhysRevB.106.155304},
	url = {https://link.aps.org/doi/10.1103/PhysRevB.106.155304}
}

@Article{Rodriguez2008,
	author={Rodr{\'i}guez, A. H.
	and Ram{\'i}rez, H. Y.},
	title={Analytical calculation of eigen-energies for lens-shaped quantum dot with finite barriers},
	journal={The European Physical Journal B},
	year={2008},
	month={Nov},
	day={01},
	volume={66},
	number={2},
	pages={235-238},
	issn={1434-6036},
	doi={10.1140/epjb/e2008-00394-3},
	url={https://doi.org/10.1140/epjb/e2008-00394-3}
}

@Article{QD-molecules-2014,
	author={Fino, Nelson R.
	and Camacho, Angela S.
	and Ram{\'i}rez, Hanz Y.},
	title={Coupling effects on photoluminescence of exciton states in asymmetric quantum dot molecules},
	journal={Nanoscale Research Letters},
	year={2014},
	month={Jun},
	day={12},
	volume={9},
	number={1},
	pages={297},
	issn={1556-276X},
	doi={10.1186/1556-276X-9-297},
	url={https://doi.org/10.1186/1556-276X-9-297}
}

@article{Knobloch2021,
author = {Knobloch, Theresia and Illarionov, Y.Y. and Ducry, Fabian and Schleich, Christian and Wachter, Stefan and Watanabe, Kenji and Taniguchi, Takashi and Mueller, Thomas and Waltl, Michael and Lanza, Mario and Vexler, Mikhail and Luisier, Mathieu and Grasser, Tibor},
year = {2021},
month = {02},
pages = {98-108},
title = {The performance limits of hexagonal boron nitride as an insulator for scaled CMOS devices based on two-dimensional materials},
volume = {4},
journal = {Nature Electronics},
doi = {10.1038/s41928-020-00529-x}
}

@Article{electrical-trapping-1,
	author={Thureja, Deepankur
	and Imamoglu, Atac
	and Smole{\'{n}}ski, Tomasz
	and Amelio, Ivan
	and Popert, Alexander
	and Chervy, Thibault
	and Lu, Xiaobo
	and Liu, Song
	and Barmak, Katayun
	and Watanabe, Kenji
	and Taniguchi, Takashi
	and Norris, David J.
	and Kroner, Martin
	and Murthy, Puneet A.},
	title={Electrically tunable quantum confinement of neutral excitons},
	journal={Nature},
	year={2022},
	month={Jun},
	day={01},
	volume={606},
	number={7913},
	pages={298-304},
	issn={1476-4687},
	doi={10.1038/s41586-022-04634-z},
	url={https://doi.org/10.1038/s41586-022-04634-z}
}

@article{electrical-trapping-2,
	author = {Jenny Hu  and Etienne Lorchat  and Xueqi Chen  and Kenji Watanabe  and Takashi Taniguchi  and Tony F. Heinz  and Puneet A. Murthy  and Thibault Chervy },
	title = {Quantum control of exciton wave functions in 2D semiconductors},
	journal = {Science Advances},
	volume = {10},
	number = {12},
	pages = {eadk6369},
	year = {2024},
	doi = {10.1126/sciadv.adk6369},
	URL = {https://www.science.org/doi/abs/10.1126/sciadv.adk6369},
	eprint = {https://www.science.org/doi/pdf/10.1126/sciadv.adk6369}}

@Article{Control-of-gap,
	author={Tebbe, David
	and Sch{\"u}tte, Marc
	and Watanabe, Kenji
	and Taniguchi, Takashi
	and Stampfer, Christoph
	and Beschoten, Bernd
	and Waldecker, Lutz},
	title={Tailoring the dielectric screening in WS2--graphene heterostructures},
	journal={npj 2D Materials and Applications},
	year={2023},
	month={Apr},
	day={08},
	volume={7},
	number={1},
	pages={29},
	issn={2397-7132},
	doi={10.1038/s41699-023-00394-0},
	url={https://doi.org/10.1038/s41699-023-00394-0}
}

@Article{Exciton-funneling-2022,
	author={Su, Haowen
	and Xu, Ding
	and Cheng, Shan-Wen
	and Li, Baichang
	and Liu, Song
	and Watanabe, Kenji
	and Taniguchi, Takashi
	and Berkelbach, Timothy C.
	and Hone, James C.
	and Delor, Milan},
	title={Dark-Exciton Driven Energy Funneling into Dielectric Inhomogeneities in Two-Dimensional Semiconductors},
	journal={Nano Letters},
	year={2022},
	month={Apr},
	day={13},
	publisher={American Chemical Society},
	volume={22},
	number={7},
	pages={2843-2850},
	issn={1530-6984},
	doi={10.1021/acs.nanolett.1c04997},
	url={https://doi.org/10.1021/acs.nanolett.1c04997}
}

@Article{nanofabrication-1,
	author ={Buchheim, Jakob and Wyss, Roman M. and Shorubalko, Ivan and Park, Hyung Gyu},
	title  ={Understanding the interaction between energetic ions and freestanding graphene towards practical 2D perforation},
	journal  ={Nanoscale},
	year  ={2016},
	volume  ={8},
	issue  ={15},
	pages  ={8345-8354},
	publisher  ={The Royal Society of Chemistry},
	doi  ={10.1039/C6NR00154H},
	url  ={http://dx.doi.org/10.1039/C6NR00154H}
}

@Article{nanofabrication-2,
	author={Danielsen, Dorte R.
	and Lyksborg-Andersen, Anton
	and Nielsen, Kirstine E. S.
	and Jessen, Bjarke S.
	and Booth, Timothy J.
	and Doan, Manh-Ha
	and Zhou, Yingqiu
	and B{\o}ggild, Peter
	and Gammelgaard, Lene},
	title={Super-Resolution Nanolithography of Two-Dimensional Materials by Anisotropic Etching},
	journal={ACS Applied Materials {\&} Interfaces},
	year={2021},
	month={Sep},
	day={08},
	publisher={American Chemical Society},
	volume={13},
	number={35},
	pages={41886-41894},
	issn={1944-8244},
	doi={10.1021/acsami.1c09923},
	url={https://doi.org/10.1021/acsami.1c09923}
}

@article{exciton-exciton-interaction,
	title = {Exciton-exciton interaction in transition metal dichalcogenide monolayers and van der Waals heterostructures},
	author = {Erkensten, Daniel and Brem, Samuel and Malic, Ermin},
	journal = {Phys. Rev. B},
	volume = {103},
	issue = {4},
	pages = {045426},
	numpages = {8},
	year = {2021},
	month = {Jan},
	publisher = {American Physical Society},
	doi = {10.1103/PhysRevB.103.045426},
	url = {https://link.aps.org/doi/10.1103/PhysRevB.103.045426}
}

@article{exciton-linewidth,
	title = {Excitonic Linewidth Approaching the Homogeneous Limit in ${\mathrm{MoS}}_{2}$-Based van der Waals Heterostructures},
	author = {Cadiz, F. and Courtade, E. and Robert, C. and Wang, G. and Shen, Y. and Cai, H. and Taniguchi, T. and Watanabe, K. and Carrere, H. and Lagarde, D. and Manca, M. and Amand, T. and Renucci, P. and Tongay, S. and Marie, X. and Urbaszek, B.},
	journal = {Phys. Rev. X},
	volume = {7},
	issue = {2},
	pages = {021026},
	numpages = {12},
	year = {2017},
	month = {May},
	publisher = {American Physical Society},
	doi = {10.1103/PhysRevX.7.021026},
	url = {https://link.aps.org/doi/10.1103/PhysRevX.7.021026}
}

\end{document}